\documentclass[seceq]{ptptex}

\newcommand{\bea}{\begin{eqnarray}}
\newcommand{\eea}{\end{eqnarray}}
\newcommand{\bi}{\begin{itemize}}
\newcommand{\ei}{\end{itemize}}

\def\s{\sigma}

\def\P{{\cal P}}

\def\bfphi{{\bf \phi}}

\def\bx{{\bf x}}
\def\bk{{\bf k}}

\def\beq{\begin{equation}}
\def\eeq{\end{equation}}

\def\P{{\cal P}}

\def\curv{\sigma}

\def\z{z}
\def\N{N}
\def\tOm{{\tilde\Omega}}

\def\r{r}
\def\ef{\varepsilon_f}
\def\tf{\tilde f}
\def\tg{\tilde g}

\title{
Primordial non-Gaussianities %
}

\author{
David \textsc{Langlois}%
}

\inst{
APC, CNRS-Universit\'e Paris 7\\
10, rue Alice Domon et L\'eonie Duquet, 75205 Paris Cedex 13, France\\
and
\\
IAP, 98bis Boulevard Arago, 75014 Paris, France
}

\abst{
This contribution gives an overview  on primordial non-Gaussianities from a theoretical perspective. After presenting a general formalism to describe nonlinear cosmological perturbations, several classes of models,  illustrated with examples, are discussed: multi-field inflation with non-standard Lagrangians, modulaton fields, curvaton fields. In the latter case, a special emphasis is put on the isocurvature perturbations, which could leave a specific signature in non-Gaussianities.
}

\begin{document}

\maketitle

\section{Introduction}

A potentially promising probe of the early Universe, which has been studied very actively in the last few years, is the non-Gaussianity of the primordial perturbations. Whereas the simplest models of inflation, based on a {\it single} field with {\it standard} kinetic term produce undetectable levels of non-Gaussianity, a significant amount of non-Gaussianity  can be produced in scenarios with i) non-standard kinetic terms; ii) multiple fields; iii) a non standard vacuum; iv) a breakdown of slow-roll evolution. 

In this contribution, after summarizing   the formalism  to describe primordial non-Gaussianities, we review three categories of models that could produce detectable non-Gaussianities.
First, we discuss inflationary models with generalized Lagrangians involving multiple scalar fields and non standard kinetic terms. Second, we consider non-Gaussianity generated by modulaton fields, or light scalar fields which are spectator fields during inflation but can affect some cosmological transition. Finally, we present some aspects of non-Gaussianity arising from curvaton models, in particular the issue of isocurvature non-Gaussianity.

\section{Nonlinear cosmological perturbations}
In this section, we  first present a geometrical description of the nonlinear cosmological perturbations. We then summarize the traditional statistical  description of primordial perturbations, which are used to relate early Universe models with cosmological observations.

\subsection{Covariant approach}
Instead of  the traditional metric-based approach~\cite{Malik:2008im}, our discussion below is based  on a more geometrical approach~\cite{Langlois:2010vx}.
Let us  consider a spacetime with metric $g_{ab}$ and some perfect fluid characterized by its energy density $\rho$, its pressure $P$ and its four-velocity $u^a$. The corresponding energy momentum-tensor is given by
\beq 
T_{ab}=\rho\,  u_a u_b+P(g_{ab}+u_au_b).
\eeq
Let us also introduce the expansion along the fluid worldlines,
\beq
 \Theta\equiv \nabla_a u^a, 
\eeq
and the integrated expansion
\beq 
 \quad N\equiv\frac{1}{3}\int d\tau \,
\Theta \;,
\eeq
where $\tau$ is the proper time defined along the fluid worldlines.  By noting that  $\Theta/3$ corresponds to the Hubble parameter $H$ in a homogeneous and isotropic  spacetime, one can interpret $\Theta/3$, in the general case, as a {\it local} Hubble parameter and $a_{\rm loc}\equiv e^N$ as a local scale factor, $N$ being the local number of e-folds. 
 
The conservation law for the energy-momentum tensor, $\nabla_a T^{a}_{\ b}=0$, 
implies~\cite{Langlois:2005ii,Langlois:2005qp} that the covector
\beq
 \zeta_a\equiv
\nabla_a N-\frac{\dot N}{\dot\rho}\nabla_a\rho=\nabla_a N+\frac{\nabla_a\rho}{3(\rho+P)}
\label{zeta_a}
\eeq
satisfies the relation
\beq
\label{dot_zeta}
\dot\zeta_a\equiv {\cal L}_u\zeta_a=
-\frac{\Theta}{3(\rho+P)}\left( \nabla_a P -
\frac{\dot P}{\dot \rho} \nabla_a\rho\right) \;,
\eeq
 where a dot denotes a Lie derivative along $u^a$, which is equivalent to an ordinary
derivative for {\it scalar} quantities (e.g. $\dot\rho\equiv u^a\nabla_a\rho$).

If $w\equiv P/\rho$ is constant,
the above covector  is a total gradient and can be written as
\beq
\zeta_a=\nabla_a\left[N+\frac{1}{3(1+w)}\ln \rho\right]\, .
\label{zeta_a_w}
\eeq
On scales larger than the Hubble radius, the above definition is
equivalent to the non-linear curvature perturbation on uniform
density hypersurfaces~\cite{Lyth:2004gb}
 \beq \zeta = \delta N -
 \int_{\bar \rho}^{\rho} H \frac{d \tilde \rho}{\dot{ \tilde \rho}}
 =
 \delta N + \frac13\int_{\bar \rho}^{\rho} \frac{d
 \tilde{\rho}}{(1+w)\tilde{ \rho}}\;.
 \eeq
  The
above equation is simply the  integrated version of
(\ref{zeta_a}).

The covector $\zeta_a$, or the corresponding scalar quantity $\zeta$,  can be defined for the global cosmological fluid or for any of the
individual cosmological fluids (this approach can also be extended to the case of interacting fluids\cite{Langlois:2006iq}).

\subsection{Power spectrum and higher order correlation functions}
We now discuss the statistical properties of the cosmological perturbation $\zeta$. 
We first define the power spectrum
\beq
\langle
 \zeta_{\bk_1} \zeta_{\bk_2} 
\rangle \equiv (2 \pi)^3
\delta^{(3)}(\bk_1+\bk_2) 
P_\zeta (k_1)\, ,
\eeq
where the Fourier modes are defined by 
\beq
\label{fourier2}
\zeta_\bk=(2\pi)^3\int d^3\bx \ e^{-i \bk\cdot\bx} \, \zeta(\bx)\, .
\eeq

Beyond the Gaussian case, the first quantity of interest is the three-point function, or its Fourier transform, called  the bispectrum and defined by 
\beq
\langle
 \zeta_{\bk_1} \zeta_{\bk_2} \zeta_{\bk_3} 
\rangle \equiv (2 \pi)^3
\delta^{(3)}(\sum_i \bk_i) 
B_\zeta (k_1,k_2,k_3)\, .
\eeq
Equivalently, one often uses the so-called $f_{\rm NL}$ parameter, which can be defined in general by
\beq
\label{bispectrum}
B_\zeta (k_1,k_2,k_3)\equiv
\frac{6}{5}
f_{\rm NL} (k_1,k_2,k_3)\left[P_\zeta(k_1)P_\zeta(k_2)+P_\zeta(k_2)P_\zeta(k_3)+P_\zeta(k_3)P_\zeta(k_1)\right]\, .
\eeq
 Similarly,  the Fourier transform of the connected four-point function defines the trispectrum, according to
\beq
\langle \zeta_{\bk_1} \zeta_{\bk_2} \zeta_{\bk_3} \zeta_{\bk_4} \rangle_{c} \equiv (2 \pi)^3
\delta^{(3)}(\sum_i \bk_i) 
T_\zeta (\bk_1, \bk_2, \bk_3, \bk_4)\, .
\eeq
The various correlation functions defined above can in principle be measured, or at least constrained, by observations of  CMB fluctuations
 or large scale structure.

In many models, the final perturbation $\zeta$ depends on the fluctuations of one or several scalar fields  generated during inflation
and  the so-called $\delta N$-formalism~\cite{Starobinsky:1986fxa, Sasaki:1995aw} provides  a powerful method  to  evaluate, at least formally,  the primordial non-Gaussianity generated on large scales
\cite{Lyth:2005fi}. The underlying idea  is to describe, on scales larger than the Hubble radius, the non-linear evolution of perturbations generated during  inflation in
terms of the perturbed expansion from an initial hypersurface
(usually taken at Hubble crossing during inflation) up to a final uniform-density
hypersurface (usually during the radiation-dominated era).
Using the Taylor expansion of the number of e-folds given as a function of the initial values of the scalar fields, 
\beq
\label{taylor}
\zeta \simeq  \sum_I N_{,I} \delta \varphi_*^I + \frac{1}{2}
\sum_{IJ} N_{,IJ} \delta \varphi_*^I \delta \varphi_*^J
\eeq
one finds~\cite{Lyth:2005fi,Seery:2005gb}, in Fourier space, 
\begin{eqnarray}
\langle \zeta_{\bk_1} \zeta_{\bk_2} \zeta_{\bk_3} \rangle &=&
\sum_{IJK} N_{,I} N_{,J} N_{,K} \langle \delta \varphi^I_{\bk_1}
\delta \varphi^J_{\bk_2} \delta \varphi^K_{\bk_3}\rangle + \nonumber
\\ &&  \frac{1}{2} \sum_{IJKL} N_{,I} N_{,J} N_{,KL} \langle \delta
\varphi^I_{\bk_1} \delta \varphi^J_{\bk_2} (\delta \varphi^K \star
\delta \varphi^L)_{\bk_3}\rangle
+{\rm perms}. \nonumber \\ &&
\label{tpf}
\end{eqnarray}
From this expression, one can distinguish two types  of  non-Gaussianity, depending on whether the first line or the second line dominates.   

In the first case, non-Gaussianities arise from the three-point function of the scalar field(s), as models with non-standard kinetic terms~\cite{Creminelli:2003iq,Seery:2005wm,Chen:2006nt}. This leads to   a specific shape of non-Gaussianity, called {\it equilateral} because the dominant contribution comes from configurations where the three wavevectors have similar length $k_1\sim k_2\sim k_3$. 

In the second case, non-Gaussianities arise from the nonlinear dependence of $N$ of the scalar field(s). Assuming  quasi-Gaussian scalar field fluctations, one finds  
that  (\ref{tpf}) leads to a bispectrum of the form (\ref{bispectrum}) with
\beq
\label{f_local}
\frac{6}{5}f_{\rm NL} = 
  \frac{
N_{I} N_{J} N^{IJ}}{( N_{K}N^K)^2}\, .
\eeq
This corresponds to another shape of non-Gaussianity, usually called {\it local} or {\it squeezed}, for which the dominant contribution comes from configurations where the three wavevectors form a squeezed triangle.

The present observational constraints~\cite{Komatsu:2010fb}  on these two main types of non-Gaussianity are
\beq
-10< f_{NL}^{\rm (local)}<74 \quad (95\% \, {\rm CL}), \qquad -214< f_{NL}^{\rm (equil)}<266 \quad (95\%\,  {\rm CL})\,.
\eeq
Note that other types on non-Gaussianity can be produced, such as the  {\it folded} shape (which peaks at $k_3 \sim k_1+k_2$) arising from a non-standard vacuum~\cite{Chen:2006nt}.

Extending  the Taylor expansion (\ref{taylor}) up to third order, one can compute in a similar way the trispectrum~\cite{Seery:2006vu}. For  local non-Gaussianity,  the trispectrum can be written in the form~\cite{Byrnes:2006vq}
\begin{eqnarray}
\label{trispectrum}
T_\zeta (\bk_1, \bk_2, \bk_3, \bk_4)&=&\tau_{\rm NL}\left[P(k_{13})P (k_3)P(k_4)+ 11 \ {\rm perms}\right]
\cr
&& +\frac{54}{25} g_{\rm NL}\left[P(k_2)P(k_3)P(k_4)+3\ {\rm perms}\right],
\end{eqnarray}
with 
 \beq
 \label{g_local}
  \tau_{\rm NL}= \frac{N_{IJ}N^{IK}N^JN_K}{(N_LN^L)^3}, \qquad
  g_{\rm NL}=\frac{25}{54}\frac{N_{IJK}N^IN^J N^K}{(N_LN^L)^3}
  \eeq
and where $k_{13}\equiv\left|\bf {k}_1+\bf {k}_3\right|$.
The present constraints on these parameters, assuming the data do not contain any isocurvature contribution, are~\cite{Smidt:2010sv}
\begin{eqnarray*}
-7.4 <10^{-5} g_{NL} < 8.2 \  (95\%\,  {\rm CL}) , \qquad  -0.6  < 10^{-4} \tau_{NL} < 3.3 \ (95\%\,  {\rm CL})\,.
\end{eqnarray*}

The next section will be devoted to models with several inflatons described by a very general Lagrangian, with multi-field DBI inflation as a concrete illustration. Later, we will consider in turn two types of scenarios leading to local type non-Gaussianity: modulatons and curvatons. 

\section{Generalized multi-field inflation}
Although it is obviously simpler to deal with a   {\it single}  inflaton, high energy physics models usually predict  the existence of many scalar fields. Multi-field inflation should thus be considered seriously. 
Moreover, many models involve kinetic terms that are not standard. It is therefore useful  to develop a formalism that is as general  as possible and try to tackle
  multi-field models   described by an
action of the form
\beq
\label{P}
S =  \int d^4 x \sqrt{-g}\left[\frac{R}{16\pi G}   +   P(X^{IJ},\phi^K)\right] 
\eeq
where $P$ is an arbitrary function of $N$ scalar fields and their associated $N (N+1)/2$  kinetic terms
\beq
\label{X}
X^{IJ}=-\frac12 \nabla_\mu \phi^I  \nabla^\mu \phi^J.
\eeq
This  includes multi-field models of the form
\beq
P(X,\phi^I)=X-V(\phi^I), \qquad X\equiv G_{IJ} X^{IJ},
\eeq
where $G_{IJ}$ is an arbitrary metric in field space,  as well as k-inflation models~\cite{ArmendarizPicon:1999rj}, which  depend on  very general single-field Lagrangians.

The expansion up to second order in the linear perturbations of the action (\ref{P}) is useful to obtain the classical equations of motion for the perturbations and to calculate the spectra of the primordial perturbations generated  during inflation.  
 Working for convenience with the scalar field perturbations 
$Q^I$ defined in the spatially flat gauge, the
 second order action can be  written in the compact form~\cite{Langlois:2008qf}
 \begin{eqnarray}
S_{(2)} &=& \frac{1}{2} \int {\rm d}t \, {\rm d}^3x \, a^3 \left[ 
\left(P_{<IJ>} + 2 P_{<MJ>,<IK>}X^{MK}\right) \dot{Q}^{I}\dot{Q}^{J}  
\right.
\cr
&& 
\qquad \left.
- P_{<IJ>} h^{ij} \partial_iQ^I \partial_jQ^J 
 - {\cal M}_{KL}Q^K Q^L 
 + 2 \,\Omega_{KI}Q^K \dot{Q}^I  \right] \qquad
 \label{S2}
\end{eqnarray}
where the mass matrix 
${\cal M}_{KL}$ 
and the mixing matrix is
$\Omega_{KI}$
can be expressed in terms of the $X^{IJ}$, of $P$ and its derivatives.

An interesting example, which combines non-standard kinetic terms with a possibly  multi-dimensional inflaton space, is multi-field DBI (Dirac-Born-Infeld) inflation. In these models,  motivated by string theory,  inflation  is due to  the motion  of a  $D3$-brane in an internal  six-dimensional compact space~\cite{st,ast}.
 The dynamics of the brane is governed by the Dirac-Born-Infeld Lagrangian, hence the name of these models. 
 
 After some appropriate reparametrizations, the Lagrangian can be written in the form
 \beq
P= -\frac{1}{f(\bfphi^I)}\left(\sqrt{ \det(\delta^{\mu}_{\nu}+f \, G_{IJ}\partial^{\mu} \phi^I \partial_{\nu} \phi^J )}-1\right) -V(\bfphi^I),
\label{DD}
\eeq
where the  potential  arises from the brane's interactions with bulk fields or other branes. 
This Lagrangian  can be written explicitly in the form (\ref{P}) upon using
\begin{eqnarray}
\label{def_explicit}
\det(\delta^{\mu}_{\nu}+f \, G_{IJ}\partial^{\mu} \phi^I \partial_{\nu} \phi^J )
&=&1-2f G_{IJ}X^{IJ}+4f^2 X^{[I}_IX_J^{J]} \cr
&&
-8f^3 X^{[I}_IX_J^{J} X_K^{K]}+16f^4 X^{[I}_IX_J^{J} X_K^{K}X_L^{L]}
\end{eqnarray}
where the field indices are lowered by the field metric $G_{IJ}$, which corresponds  to the metric of the internal compact space,  and the brackets denote antisymmetrization over the indices. Note that the dilaton and the various form fields are ignored in the Lagrangian (\ref{DD}),  but they can also be included in the analysis of the cosmological perturbations generated by these models~\cite{Langlois:2009ej}.

  As in other multi-field inflationary models, it can be convenient~\cite{Gordon:2000hv} to decompose the perturbations into a so-called (instantaneous) adiabatic mode, along the background velocity in field space, and  (instantaneous) entropic modes, which are orthogonal to the adiabatic direction.
Focussing, for simplicity,  on the two-field case, where   there is a single entropic degree of freedom, one can  decompose 
the scalar field perturbations as
\beq
\label{decomposition}
Q^I=Q_\s e_\sigma^I+Q_s e^I_s\,,
\eeq
where the adiabatic vector $e^I_\sigma$ and entropic vector $e^I_s$ are normalized (via the field space metric $G_{IJ}$). 
The perturbations generated during inflation can then be determined by using the standard techniques, which gives~\cite{Langlois:2008wt}
\beq
{\cal P}_{Q_\s*}\simeq\frac{H^2}{4\pi^2 }, \qquad  {\cal P}_{Q_s*}\simeq\frac{H^2}{4\pi^2 c_s^2},
\eeq
(the subscript $*$ here indicates that the corresponding quantity is evaluated at sound horizon crossing $k c_s=aH$). For small $c_s$, the entropic modes are thus {\it amplified} with respect to the adiabatic modes.

Since we are in a multi-field scenario,  the curvature perturbation can evolve after sound horizon crossing, and the spectrum of the final curvature perturbation, which is probed by cosmological observations, can be formally written as
\beq
{\cal P}_{\cal R}=(1+T_{{\cal R} {\cal S}}^2) {\cal P}_{\cal R_{*}}=(1+T_{{\cal R} {\cal S}}^2) \frac{H^4}{4\pi^2\dot\sigma^2}=(1+T_{{\cal R} {\cal S}}^2) \frac{H^2}{8\pi^2\, \epsilon \, c_s}
\label{observed-spectrum}
\eeq
 where $T_{{\cal R} {\cal S}}$ quantifies the transfer  from the entropic into the adiabatic modes. 
  
Let us now discuss non-Gaussianities in multi-field DBI inflation. The three-point correlation functions of the scalar fields can be computed from the third order action, which is given, in the small sound speed limit, by~\cite{Langlois:2008wt,Langlois:2008qf}
\begin{eqnarray}
S_{(3)}&=&\int {\rm d}t\, {\rm d}^3x\,  \left\{ \frac{a^3}{2 c_s^5 \dot \s}\left[(\dot Q_{\s} )^3+c_s^2 \dot Q_{\s}  (\dot Q_{s} )^2\right]  \right.
 \cr
 && \left. 
 - \frac{a}{2 c_s^3 \dot \s}\left[ \dot Q_{\s} (\partial  Q_{\s} )^2 -c_s^2 \dot{Q_{\s} }(\partial Q_s)^2+2 c_s^2 \dot {Q_s}\partial Q_{\s} \partial Q_s)\right]
 \right\}\, .
 \label{S3}
\end{eqnarray}
The contribution from the scalar field three-point functions  to the coefficient $f_{\rm NL}$ is  found to be given by
\beq
f_{NL}^{(3)}=-\frac{35}{108 \, c_s^2 \, (1+T^2_{{\cal R} {\cal S}}) } \,,
\label{f_NL3}
\eeq 
which is similar to the single-field DBI result,  but with a suppression due to  the transfer between the entropic and adiabatic modes.  

Interestingly, multi-field DBI inflation could also produce a local non-Gaussianity in addition to the equilateral one~\cite{RenauxPetel:2009sj}. Finally, let us mention
 that the trispectrum in  multi-field DBI inflation has also been computed~\cite{Mizuno:2009mv}.

\section{Modulatons}
\def\mod{\sigma}
Significant non-Gaussianity can arise when a cosmological transition in the history of the Universe depends on some light scalar field, which has previoulsy acquired some fluctuations during the inflationary phase. Consequently, in different regions of the Universe where the value of the scalar field is slightly different, the cosmological transition and the subsequent cosmological evolution will differ. 
In this way, the fluctuations of the scalar field, which we will call  a {\it modulaton}, are converted into curvature fluctuations.

\subsection{Modulated reheating}
A typical  example is the {\it modulated reheating} scenario~\cite{Dvali:2003em,Kofman:2003nx}  where the decay rate of the inflaton, $\Gamma$,  depends on a modulaton $\mod$.

A simple way to compute the curvature perturbation is to calculate  the number of e-folds between some initial time $t_i$ during inflation, when the scale of interest crossed out the Hubble radius, and some final time $t_f$. For simplicity, let us  assume that,  just after the end of inflation at time $t_e$, the inflaton behaves like pressureless matter (as is the case for a quadratic potential) until it decays instantaneously at the time $t_d$
 characterized by $H_d=\Gamma$. 
 At the decay, the energy density is thus $\rho_d=\rho_e\exp[-3(N_d-N_e)]$ and is transferred into radiation, so that, at a subsequent  time $t_f$, one gets
 \beq
 \label{rho_f_modulaton}
 \rho_f=\rho_de^{-4(N_f-N_d)}=\rho_e e^{-3(N_f-N_e)-(N_f-N_d)}.
 \eeq
 Using the relation  $\Gamma=H_d=H_f\exp[2(N_f-N_d)]$ to eliminate $(N_f-N_d)$ in (\ref{rho_f_modulaton}), we finally obtain 
 \beq 
 N_f=N_e-\frac13\ln\frac{\rho_f}{\rho_e}-\frac16\ln\frac{\Gamma}{H_f}\, .
 \eeq
 This implies that the nonlinear  curvature perturbation can be simply expressed as 
 \beq
 \label{zeta_modulaton}
 \zeta=\zeta_{\rm inf}-\frac16\ln\left(\frac{\Gamma(\mod)}{\bar\Gamma}\right)\,,
\eeq
where $\zeta_{\rm inf}$ represents the contribution to the curvature perturbation from the inflaton fluctuations. 
At linear level, this  leads to  the curvature power spectrum  
\beq
\P_\zeta=\P_{\zeta_{\rm inf}}+\frac{1}{36}\left(\frac{\Gamma_{,\mod}}{\Gamma}\right)^2\P_{\delta\sigma_*}=\P_{\zeta_{\rm inf}}+\frac{1}{36}\left(\frac{\Gamma_{,\mod}}{\Gamma}\right)^2\left(\frac{H_*}{2\pi}\right)^2\, .
\eeq
By expanding (\ref{zeta_modulaton}) up to  second and third orders in $\delta\mod_*$, one can easily determine the bispectrum and trispectrum for the 
curvature perturbation. Using the expressions (\ref{f_local}) and (\ref{g_local}), one finds that  the associated nonlinear parameters are given by 
\beq
f_{\rm NL}=5\left(1-\frac{\Gamma \Gamma''}{\Gamma^{\prime\, 2}}\right)\Xi^2,
\eeq 
and 
\beq
\tau_{\rm NL}=\frac{36}{25}\, f_{\rm NL}^2 \, \Xi^{-1}\,,
 \quad g_{\rm NL}=\frac{50}{3}\left(2-3\frac{\Gamma \Gamma''}{\Gamma^{\prime\, 2}}+\frac{\Gamma^2\Gamma'''}{\Gamma^{\prime\, 3}}\right)\Xi^3\,.
\eeq
where $\Xi=1-\left(\P_{\zeta_{\rm inf}}/\P_{\zeta}\right)$ represents the fraction of the curvature power spectrum due to the modulaton. 

\subsection{Modulated trapping}
One can also envisage the possibility that a  modulaton field  affects the cosmological evolution {\it during} inflation.  This is the case   in the {\it modulated trapping} scenario~\cite{Langlois:2009jp}, which relies on the resonant production of particles during inflation~\cite{Chung:1999ve}. In this model, the  inflaton $\phi$ is coupled to other fields, for example to some fermions $\psi$ via  with the interaction Lagrangian ${\cal L}_{\rm int}=\lambda\phi\,\bar{\psi}\psi$.

If, during inflaton, the effective mass of $\psi$, $m_{\rm eff}=m-\lambda\phi$ becomes zero,  this triggers a burst of production of these particles. Even if these particles are  quickly diluted by the exansion, their backreation will  affect the evolution of the inflaton, governed by the equation of motion
\begin{equation}\label{kg}
\ddot{\phi}+3\,H\,\dot\phi+V'\left(\phi\right)=\N\,\lambda\langle \bar\psi\psi\rangle =\lambda  n_*\left(\frac{a}{a_*}\right)^{-3}\Theta(t-t_*).
\end{equation}
Indeed, the term on the right hand side induces  a temporary slow-down of the inflaton, which leads to a slightly  longer phase of inflation. This brief trapping of the inflaton thus manifests itself as  an increment of   the number of e-folds until the end of inflation:
\beq
\label{N_trapping}
N=N_{\rm std}(\phi) + \Delta N_{\rm trapping}
\eeq

Let now assume that this trapping depends on some modulaton, for example via the coupling between the inflaton and the particles, and occurs well after the fluctuations of the modulaton (on observable cosmological scales) have been generated. This is in contrast with other scenarios~\cite{Chung:1999ve,Elgaroy:2003hp,Romano:2008rr,Barnaby:2009mc}  where the trapping occurs approximately when cosmological scales exit the Hubble radius, which leads to special features in the  CMB spectrum as well as specific non-Gaussianity~\cite{Barnaby:2010ke}. 

Using the Taylor expansion of (\ref{N_trapping}), where only the second term $\Delta N$ depends on the modulaton $\mod$:
\beq
\zeta=\delta N= \frac{d N_{\rm slow-roll}}{d\phi}\delta\phi+\dots+ \Delta N_{,\mod}\delta\mod+\frac12\Delta N_{,\mod\mod}\delta\mod^2
+\frac16\Delta N_{,\mod\mod\mod}\delta\mod^3\, 
\eeq
(higher order derivatives with respect to the inflaton are ignored, because they give  negligible non-Gaussianities), one can compute the power spectrum and non-Gaussianity of the  curvature perturbation, generated by the modulated trapping scenario~\cite{Langlois:2009jp}.
  According to (\ref{f_local}), the corresponding 
non-linearity parameter for the bispectrum is given by
\beq
\label{f_NL}
\frac{6}{5}f_{\rm NL} = 
  \frac{
(\Delta N_{,\mod})^2 \Delta N_{,\mod\mod}}{\left({N^{\rm sr}_{,\phi}}^2+(\Delta N_{,\mod})^2\right)^2}
=\left(\frac{\P_\zeta^{\rm trapping}}{\P_\zeta}\right)^2\frac{\Delta N_{,\mod\mod}}{(\Delta N_{,\mod})^2}
=\Xi^2\, \frac{\Delta N_{,\mod\mod}}{(\Delta N_{,\mod})^2}
.
\eeq
Similarly, the nonlinear coefficients of the trispectrum  are 
\beq
\label{g_NL}
\tau_{\rm NL}
= \frac{(\Delta N_{,\mod\mod})^2}{(\Delta N_{,\mod})^4}\ \Xi^3=\frac{36}{25\Xi}f_{\rm NL}^2\,, \qquad
g_{\rm NL}=\frac{25}{54}\frac{\Delta N_{,\mod\mod\mod}}{(\Delta N_{,\mod})^3}\ \Xi^3
\eeq
 
The most interesting situation occurs   when   the coupling $\lambda$ depends directly on the modulaton, which leads to the nonlinear parameters
\beq
f_{\rm NL}=\frac{1}{2e \beta}\left(3+2\frac{\lambda\lambda''}{\lambda'^2}\right)\Xi^2,\quad  g_{\rm NL}=\frac{1}{2e^2\beta^2}\left[1+6\frac{\lambda\lambda''}{\lambda'^2}+4\frac{\lambda^2\lambda'''}{3\lambda'^3}\right]\, \Xi^3\, ,
\eeq
where $\beta\equiv {\rm Max}(\Delta\dot\phi)/\vert\dot\phi_*\vert$ cannot exceed $1$.

If $\lambda$ depends only linearly on $\mod$, the first  expression reduces to $f_{\rm NL}=3\Xi^2/(2e \beta)$, which shows  that it is quite easy to obtain a detectable level of non-Gaussianity in this scenario: for example, one gets $f_{NL}\simeq 55$ with $\beta=0.01$ and $\Xi=1$.  Moreover, there is a specific relation between $\tau_{\rm NL}$ and $g_{\rm NL}$ which could be confronted with observations if these quantities can be measured, and thus distinguish this scenario from other scenarios leading to different relations between the nonlinear coefficients~\cite{Suyama:2010uj}.

If the future cosmological data point to the existence of a significant amount of {\it local} non-Gaussianity, the modulated trapping scenario would thus represent a viable model, together with the modulated reheating or the curvaton scenario which we examine in the next section.

\section{Curvatons and isocurvature perturbations}
\def\i{{\rm inf}}
\def\SG{{\hat S}}
\def\zetarad{\zeta_{\rm r}}
\def\f{f_c}

The last example that  we consider in this contribution is  the curvaton scenario~\cite{curvaton}, or more precisely in the mixed curvaton and inflaton version~\cite{Langlois:2004nn} where the inflaton fluctuations are also taken into account. 
 The curvaton is a weakly coupled scalar field,
$\curv$,
which is light relative to the Hubble
rate during inflation, and hence acquires Gaussian fluctuations with an almost
scale-invariant spectrum.  After inflation
the Hubble rate drops and eventually the curvaton becomes
non-relativistic so that its energy density grows with respect to that of  
radiation, until  it decays. 

Many aspects of the curvaton scenario have been studied in the literature. Here, we wish to focus our attention on {\it isocurvature} perturbations that can be generated in this type of scenario, and their non-Gaussianity. Isocurvature non-Gaussianity, which has been investigated recently in several works~\cite{Kawasaki:2008sn,Langlois:2008vk,Kawasaki:2008pa,Hikage:2008sk,Kawakami:2009iu,Langlois:2010dz,Langlois:2010fe}, could indeed be distinguished from the usual adiabatic non-Gaussianity and thus open a new window on the early Universe, if ever detected. 

As a preliminary step we present a general formalism that computes systematically the evolution of the  nonlinear perturbations of various fluids through a decay transition. We then apply this formalism to a scenario with a curvaton fluid, radiation and cold dark matter CDM and compute the adiabatic and isocurvature perturbations, up to third order. 

\subsection{Evolution of the perturbations due to the decay of some species}
We now consider a very general setting where several cosmological fluids coexist, each of them characterized by the nonlinear curvature perturbation
\beq
\label{defzeta}
\zeta_A=\delta N+\frac{1}{3(1+w_A)} \ln\frac{\rho_A}{\bar\rho_A}\,,
\eeq
as follows from the definition  (\ref{zeta_a_w}). 
We wish to compute the curvature perturbations after the decay of one these fluids, denoted $\curv$, which will later correspond to the curvaton.  

In the sudden decay approximation, which we adopt here, the decay  takes place on the  hypersurface characterized by  $
H_{\rm d}=\Gamma_\curv$, 
where $H_{\rm d}$ is the Hubble parameter at the decay and $\Gamma_\curv$ is 
the decay rate of $\curv$.
Since $H$ depends only on  the {\it total} energy density, the decay hypersurface is a hypersurface of uniform total energy density, with $\delta \N_{\rm d}=\zeta$, where $\zeta$ is the global curvature perturbation. 
The equality between the sum of all energy densities,  before the decay and  after the decay, thus reads
\beq
\label{bilan_decay}
\sum_A\bar{\rho}_{A-}e^{3(1+w_A)(\zeta_{A-}-\zeta)}=\bar{\rho}_{\rm decay}=\sum_B\bar{\rho}_{B+}e^{3(1+w_B)(\zeta_{B+}-\zeta)},
\eeq
where the subscripts $-$ and $+$ denote quantities defined, respectively, {\it before} and  {\it after} the transition. 
In the above formula, we have used the non-linear energy densities of the individual fluids, which can be expressed in terms of their curvature perturbation $\zeta_A$ by inverting the expression (\ref{defzeta}).

Expanding the first equality  in (\ref{bilan_decay}) up to third order, one finds
\beq
\label{zeta_3}
\zeta=\sum_A\lambda_A\, \left[\zeta_{A-}+\frac{\beta_A}{2}\left(\zeta_{A-}-\zeta\right)^2
+\frac{\beta_A^2}{6}\left(\zeta_{A-}-\zeta\right)^3
\right] \, ,
\eeq
with  the coefficients
\beq
\label{lambda_A}
 \beta_A\equiv 3(1+w_A), \quad \lambda_A\equiv\frac{\tOm_A}{\tOm}, \quad \tOm_A\equiv (1+w_A)\,  \Omega_A, \quad \tOm\equiv\sum_A\tOm_A\, ,
\eeq
where the abundance parameters are defined just before the decay: $\Omega_A\equiv {\bar\rho_{A-}}/\bar\rho_{\rm decay}$.
Note that, although the global perturbation $\zeta$ appears on both sides of  (\ref{zeta_3}), this relation can be used iteratively in order to determine, order by order, the expression of $\zeta$ in terms of all the $\zeta_{A-}$, up to third order.

Just after the decay, the energy density of  the  fluid $\curv$ is transferred into the one or several of the remaining fluids. Introducing the 
 relative branching ratios $\gamma_{A\curv}$, this means that 
 the energy density for any species $A$, just  after the decay of $\curv$,  
is simply given by
\begin{equation}
\rho_{A+} = \rho_{A-} + \gamma_{A\curv} \rho_\sigma.
\end{equation}
This relation, which is fully non-linear, can be reexpressed, upon using (\ref{defzeta}), in the form 
\beq
\label{bilan_A}
e^{\beta_A(\zeta_{A+}-\zeta)}=(1-f_A)e^{\beta_A(\zeta_{A-}-\zeta)}+f_Ae^{\beta_\sigma(\zeta_{\sigma -}-\zeta)} \, ,
\eeq
where the parameter
\beq
f_A\equiv \frac{\gamma_{A\sigma}\Omega_\sigma}{\Omega_A+\gamma_{A\sigma}\Omega_\sigma}\, 
\eeq
represents the fraction of the fluid $A$ that has been created by the decay. 

Expanding  (\ref{bilan_A}) up to third order, and using (\ref{zeta_3}), one gets
\begin{eqnarray}
\label{zetaA+}
\zeta_{A+}&=&\sum_B T_{A}^{\ B}\left[\zeta_{B-}+\frac{\beta_B}{2}\left(\zeta_{B-}-\zeta\right)^2
+\frac{\beta_B^2}{6}\left(\zeta_{B-}-\zeta\right)^3
\right] 
\cr
&& -\frac{\beta_A}{2}\left(\zeta_{A+}-\zeta\right)^2
-\frac{\beta_A^2}{6}\left(\zeta_{A+}-\zeta\right)^3
\end{eqnarray}
with the coefficients
\begin{eqnarray}
T_{A}^{\ A}&=& f_A\left(1-\frac{\beta_\curv}{\beta_A}\right)\lambda_A+(1-f_A)
\label{T1}
\\
T_{A}^{\ \curv}&=& f_A\left(1-\frac{\beta_\curv}{\beta_A}\right)\lambda_\curv+f_A\frac{\beta_\curv}{\beta_A}
\label{T2}
\\
T_{A}^{\ C}&=& f_A\left(1-\frac{\beta_\curv}{\beta_A}\right)\lambda_C\, , \quad C\neq A,\curv\, .
\label{T3}
\end{eqnarray}
Finally, after using (\ref{zeta_3}) again, one finds that (\ref{zetaA+}) yields the full expression, up to third order, of all the post-decay curvature perturbations in terms of the pre-decay curvature perturbations~\cite{Langlois:2010dz,Langlois:2010fe}:
\beq
\label{zeta_3rdorder}
\zeta_{A+}=\sum_BT_A^{\ B}\zeta_{B-}+\sum_{B, C}U_A^{BC}\zeta_{B-}\zeta_{C-}+
\sum_{B, C, D}V_A^{BCD}\zeta_{B-}\zeta_{C-} \zeta_{D-},
\eeq
with 
\begin{eqnarray*}
U_A^{BC}&\equiv&\frac12 \left[\sum_E\beta_E T_A^E(\delta_{EB}-\lambda_B)(\delta_{EC}-\lambda_C)-\beta_A(T_{AB}-\lambda_B)(T_{AC}-\lambda_C)
\right].
\end{eqnarray*}
and
\begin{eqnarray*}
V_A^{BCD}&\equiv&-\frac12 \sum_{E,F}\beta_E T_{AE}(\delta_{EB}-\lambda_B)\lambda_F\beta_F (\delta_{FC}-\lambda_C) (\delta_{FD}-\lambda_D)
\cr
&& +\frac16 \sum_E\beta_E^2 T_{AE}(\delta_{EB}-\lambda_B)(\delta_{EC}-\lambda_C)(\delta_{ED}-\lambda_D)
\cr
&&
-\beta_A(T_{AB}-\lambda_B)\left[U_A^{CD}-\frac12\sum_E\beta_E\lambda_E (\delta_{EC}-\lambda_C)(\delta_{ED}-\lambda_D)\right]
\cr
&&
-\frac16\beta_A^2(T_{AB}-\lambda_B)(T_{AC}-\lambda_C)(T_{AD}-\lambda_D)\, .
\end{eqnarray*}
The above expression thus provides a systematic computation of the post-decay curvature perturbations for all fluids in a very general setting. For scenarios with several decay transitions, the perturbations can be obtained by combining the various expressions of the type (\ref{zeta_3rdorder}) for each transition. 

\subsection{Mixed curvaton and inflaton scenario}

We now apply the general formalism presented above to  a scenario involving a curvaton $\curv$, behaving as a pressureless fluid,  in addition to radiation ($r$)   and CDM ($c$), which can lead to isocurvature perturbations~\cite{Moroi:2002rd,Lyth:2003ip}. The formula (\ref{zeta_3rdorder}) allows us to compute,  in terms of the pre-decay perturbations,  the perturbations $\zeta_r$ and $\zeta_c$ after the decay,  or equivalently the adiabatic perturbation, which coincides with $\zeta_r$ deep in the radiation era, and the CDM isocurvature perturbation
\beq
S_c= 3(\zeta_c-\zeta_r).
\eeq

For simplicity, we restrict our analysis to the situation where 
\beq
\zeta_{c-}=\zeta_{r-}\equiv\zeta_\i\,,
\eeq
by assuming that the CDM and radiation perturbations, {\it before} the curvaton decay, depend only on the inflaton fluctuations. 

The curvaton fluid isocurvature perturbation before the decay, 
$S_\curv$, can be easily related to the curvaton {\it field} fluctuations in the case of a quadradic potential. 
Indeed,  writing the (non-linear) energy density of  the oscillating curvaton defined on  the spatially flat
hypersurfaces, characterized by $\delta N=\zeta_r$ when the curvaton is still  subdominant:
\beq
\rho_\curv = m^2 \curv^2=m^2 \left( \bar\curv+\delta\curv \right)^2= \bar\rho_\curv e^{3(\zeta_\curv-\zeta_r)}=\bar\rho_\curv e^{S_\curv}\, ,
\eeq
leads to the relation
 \beq
 e^{S_\curv}
 =
 \left( 1+\frac{\delta\curv}{\bar\curv} \right)^2 \,. \label{rhorhobarcurv} 
 \end{equation}
Expanding this expression  up to  third
order,  and using the conservation of $\delta\curv/\curv$ in a quadratic potential, we obtain
 \beq
 \label{S_G}
S_\curv= \SG-\frac14 \SG^2+\frac{1}{12}\SG^3\, ,
\eeq
where the quantity 
 \beq
 \SG \equiv 2\frac{\delta\curv_*}{\bar\curv_*}\, 
 \eeq
is Gaussian.

 Using the general expressions (\ref{zeta_3rdorder}), one finds that 
 the primordial curvature perturbation is given by
\beq
\label{zetarad}
\zetarad =\zeta_\i+\z_1 \SG+\frac12 \z_2 \SG^2+\frac16 \z_3 \SG^3\, ,
\eeq
with 
\begin{eqnarray}
 z_1&=&  \frac{\r}{3}\,, \qquad  
 z_2=\frac{\r}{18}\left(3 -8\r +\frac{4r}{\xi} -2\frac{r^2}{\xi^2} \right), 
\\
z_3&=&\frac{\r^2}{54}\left(\frac{6 r^3}{\xi ^4}+\frac{24 r^2}{\xi ^2}-\frac{4 r^2}{\xi ^3}-\frac{48 r}{\xi }-\frac{15
   r}{\xi ^2}+64 r+\frac{18}{\xi }-36\right) \,,
\end{eqnarray}
where the parameter
\beq
\xi \equiv 
 \frac{\gamma_{r \, \curv}}{1-(1- \gamma_{r \, \curv}) \Omega_\curv}
\eeq
can be interpreted as  the  efficiency of the energy transfer from the curvaton into radiation ($\xi=1$ if the curvaton decays only into radiation, i.e. $\gamma_{r\sigma}=1$), 
and 
\beq
\r\equiv \xi \, {\tilde r}\,,\qquad {\tilde r}=  \frac{3\Omega_\curv}{4-\Omega_\curv}\, .
\eeq
For the primordial isocurvature perturbation, one obtains
\beq
\label{S_c}
S_c=s_1 \SG+\frac12 s_2 \SG^2+\frac16 s_3 \SG^3\, ,
\eeq
with 
\begin{eqnarray}
 s_1&=&\f-r, \qquad s_2=\frac{1}{6}\left(3 \f (1-2\f)+\frac{2 r^3}{\xi ^2}-\frac{4 r^2}{\xi }+8 r^2-3 r\right)\, ,
\\
s_3&=&-\frac12\f^2(3-4\f)-\frac{\r^2}{18}\left(\frac{6 r^3}{\xi ^4}+\frac{24 r^2}{\xi ^2}-\frac{4 r^2}{\xi ^3}-\frac{48 r}{\xi }-\frac{15
   r}{\xi ^2}+64 r+\frac{18}{\xi }-36\right).\quad \qquad
\end{eqnarray}
From these expressions, one can determine the power spectrum, the bispectrum and the trispectrum,
which can be probed by observations.

The power spectrum for the total curvature perturbation follows from the linear part and  is given by
 \beq
 \label{finalzetar}
  {\cal P}_{\zetarad}={\cal P}_{\zeta_\i}+\frac{r^2}{9}{\cal P}_{\SG} = \Xi^{-1}\frac{r^2}{9}{\cal P}_{\SG},
 \eeq
where  $\Xi$ represents the fraction of the power spectrum due to the curvaton contribution. 
The power spectrum for the isocurvature fluctuations is, 
according to Eq.~\eqref{S_c}, 
\begin{equation}
{\cal P}_{S_c} = (f_c - r)^2 {\cal P}_{\SG}.
\end{equation}
Both curvature and isocurvature perturbations depend on the curvaton fluctuations and are  therefore correlated, 
 with  the correlation 
coefficient:
\begin{equation}
\label{corr_ad_is}
{\cal C}
= \frac{ {\cal P}_{S_c, \zeta_r} }{\sqrt{ {\cal P}_{S_c} {\cal P}_{\zeta_r}}} =\ef \sqrt{\Xi}, \qquad 
\ef\equiv {\rm sgn} (f_c -r) .
\end{equation}
In the pure curvaton limit ($\Xi\simeq 1$), adiabatic and isocurvature perturbations are either fully correlated, if $\ef>0$, or fully anti-correlated, if $\ef<0$.
In the opposite limit ($\Xi\ll 1$), the correlation vanishes. 
For intermediate values of  $\Xi$, the correlation is only partial, as can also be  obtained  in multifield inflation
\cite{Langlois:1999dw}.

The isocurvature-to-adiabatic ratio
\begin{equation}
\label{eq:alpha}
\alpha = \frac{{\cal P}_{S_c}}{{\cal P}_{\zeta_r}} = 9\left(1-\frac{f_c}{ r}\right)^2 \,\Xi \, ,
\end{equation}
is strongly constrained by cosmological observations, the precise limits depending on the assumed level of correlation between the isocurvature and adiabatic perturbations (since the impact of isocurvature perturbations on the observable power spectrum depends crucially on this correlation~\cite{Langlois:2000ar}). 
In terms of  the parameter $a\equiv \alpha/(1+\alpha)$, the limits (based on WMAP+BAO+SN data) given in  \cite{Komatsu:2010fb} 
are\beq
a_0<0.064\quad (95 \% {\rm CL}), \qquad a_{1}< 0.0037 \quad (95 \% {\rm CL})\,,
\eeq
respectively for the uncorrelated case ($\Xi=0$) and for the fully correlated case ($\Xi=1$).
According to (\ref{eq:alpha}), the observational constraint $\alpha \ll 1$ can be satisfied if $|\f-r|\ll r$ (which includes the case   $\f=1$ with  $r\simeq  1$)
or if  $\Xi \ll 1$, i.e. the curvaton contribution to the observed power spectrum is very small.

\subsection{Adiabatic and isocurvature non-Gaussianities}
Since we now deal with two observable quantities, namely adiabatic and isocurvature perturbations, the definition of the bispectrum can be extended to include both types of perturbations.
In our particular case, where  there is only one  degree of freedom, $\SG$, at the nonlinear level, one can show that the generalized bispectra (with indices  $I=\{\zeta, S\}$)  are of the form~ \cite{Langlois:2010dz}
\beq
 B^{IJK}(k_1, k_2, k_3)=
 b_{NL}^{I, JK}  P_\SG(k_2) P_\SG(k_3)+b_{NL}^{J, KI}  P_\SG(k_1) P_\SG(k_3)+b_{NL}^{K, IJ}   P_\SG(k_1)P_\SG(k_2)\qquad 
 \eeq
 with 
 \beq
b_{NL}^{I, JK} \equiv N^I_{(2)} N^J_{(1)} N^K_{(1)},  
\eeq
where $N^\zeta_{(2)}=z_2$, $N^S_{(2)}=s_2$, $N^\zeta_{(1)}=z_1$, $N^S_{(1)}=s_1$, respectively.

Recalling that the usual, purely adiabatic, $f_{NL}$ is proportional to the bispectrum of $\zeta$ divided by the square of the power spectrum, one defines the analogs of $f_{NL}$  by dividing  the coefficients $b_{NL}^{I, JK}$ by the square  of the ratio $P_\zeta/P_{\SG}=z_1^2/\Xi$, i.e.
\beq
\tf_{NL}^{I,JK}\equiv \frac65 f_{NL}^{I, JK} \equiv \frac{\Xi^2}{z_1^4} \, b_{NL}^{I, JK} \, .
\eeq
Taking into account the fact that the last two indices can be permuted, this leads to   six different coefficients,  
explicitly given by the expressions
\begin{eqnarray}
\tf_{NL}^{\zeta, \zeta\zeta}&=&\frac{z_2}{z_1^2}\, \Xi^2,  \quad \tf_{NL}^{\zeta, \zeta S}=\frac{s_1 z_2}{z_1^3}\, \Xi^2,  \quad  \tf_{NL}^{S, \zeta\zeta}=
\frac{s_2}{z_1^2}\, \Xi^2,
\\
\tf_{NL}^{\zeta, S S}&=&\frac{s_1^2 z_2}{z_1^4}\, \Xi^2,  \quad \tf_{NL}^{S, \zeta S}= \frac{s_1 s_2}{z_1^3}\, \Xi^2,  \quad \tf_{NL}^{S, SS}= \frac{s_1^2 s_2}{z_1^4}\, \Xi^2\, . 
\end{eqnarray}

The same analysis applies to the  trispectra that  combine adiabatic and isocurvature perturbations, leading to  the generalized parameters~\cite{Langlois:2010fe}
\beq
\tau_{NL}^{IJ,KL} \equiv\frac{N^I_{(2)} N^J_{(2)}N^K_{(1)}N^L_{(1)}}{z_1^6}\Xi^3\, , \quad \tg_{NL}^{I,JKL}\equiv \frac{54}{25}g_{NL}^{I,JKL}\equiv \frac{N^I_{(3)} N^J_{(1)}N^K_{(1)} N^L_{(1)}}{z_1^6}\Xi^3 , 
\eeq
where $N^\zeta_{(3)}=z_3$ and  $N^S_{(3)}=s_3$.
Taking into account  the symmetries under permutations of the indices, one finds, for two observables ($I=\{\zeta, S\}$), 9 different parameters $\tau_{NL}^{IJ,KL}$ and 8 parameters $\tg_{NL}^{I,JKL}$.

An interesting question is whether one can find significant non-Gaussianities, while satisfying the bound on the isocurvature spectrum. As mentioned earlier, this isocurvature constraint can be satisfied with $\Xi\simeq 1$ if $\f$ and $\r$ are sufficiently close. In this case, one finds that the purely adiabatic non-Gaussianity dominates. But in the alternative situation where $\Xi\ll 1$, one finds that, with respect to the purely adiabatic non-Gaussianity, the  purely and mixed  isocurvature ones  are either enhanced by  constant factors, if $\f\ll r\ll 1$, or much more strongly enhanced with powers of $(3\f/r)$, if $\r\ll \f \ll 1$.

\section{Conclusions}
As this contribution has tried to illustrate with a few explicit examples, the detection of primordial non-Gaussianities would have dramatic consequences.

First, since the simplest inflationary models, based on a single field in slow roll, predict a negligible amount of primordial non-Gaussianities, these models 
would have to be replaced with more elaborate models, involving several scalar fields, non standard kinetic terms or other features. 

Second, since the measurement of non-Gaussianities contains potentially a lot of information, in particular concerning their shape, one could hope to discriminate between different categories of models, which would otherwise appear degenerate in their predictions of the  power spectrum. 

Even the simplest shape of non-Gaussianity, the local shape, can hide surprisingly rich variations if perturbations are generated by several scalar fields and if isocurvature perturbations survive. In such situation, purely adiabatic, purely isocurvature and mixed non-Gaussianities could coexist and the hierarchy between their  amplitudes would provide invaluable information.  
Since isocurvature fluctuations are usually associated 
with the generation of dark matter and baryon asymmetry in the Universe, non-Gaussianity from isocurvature fluctuations, if  detected in the future,  would give us a lot of insight into
the nature of dark matter, the mechanism of baryogenesis, and therefore into  high energy physics.

\section*{Acknowledgements}
I  am grateful to  my collaborators (Angela Lepidi, S\'ebastien Renaux-Petel, Lorenzo Sorbo, Daniele Steer, Tomo Takahashi, Takahiro Tanaka, Filippo Vernizzi, David Wands) for  our works on non-Gaussianity on which this contribution is based. I would also like to thank the organizers of the YKIS2010 for their kind invitation and the warm hospitality at the Yukawa Institute.

%


\begin{thebibliography}{99}



\bibitem{Malik:2008im}
  K.~A.~Malik, D.~Wands,
  Phys.\ Rept.\  {\bf 475}, 1-51 (2009).
  [arXiv:0809.4944 [astro-ph]].
  
\bibitem{Langlois:2010vx}
  D.~Langlois and F.~Vernizzi,
  Class.\ Quant.\ Grav.\  {\bf 27}, 124007 (2010)
  [arXiv:1003.3270 [astro-ph.CO]].

  \bibitem{Langlois:2005ii}
  D.~Langlois and F.~Vernizzi,
  Phys.\ Rev.\ Lett.\  {\bf 95}, 091303 (2005)
  [arXiv:astro-ph/0503416].

\bibitem{Langlois:2005qp}
  D.~Langlois and F.~Vernizzi,
  Phys.\ Rev.\ D {\bf 72}, 103501 (2005)
  [arXiv:astro-ph/0509078].


\bibitem{Lyth:2004gb}
  D.~H.~Lyth, K.~A.~Malik and M.~Sasaki,
  JCAP {\bf 0505}, 004 (2005)
  [arXiv:astro-ph/0411220].


\bibitem{Langlois:2006iq}
  D.~Langlois, F.~Vernizzi,
  JCAP {\bf 0602}, 014 (2006).
  [astro-ph/0601271].
  
\bibitem{Starobinsky:1986fxa}
  A.~A.~Starobinsky,
  JETP Lett.\  {\bf 42}, 152-155 (1985).
  
  
\bibitem{Sasaki:1995aw}
  M.~Sasaki and E.~D.~Stewart,
  Prog.\ Theor.\ Phys.\  {\bf 95}, 71 (1996)
  [arXiv:astro-ph/9507001];

\bibitem{Lyth:2005fi}
  D.~H.~Lyth and Y.~Rodriguez,
  Phys.\ Rev.\ Lett.\  {\bf 95}, 121302 (2005)
  [arXiv:astro-ph/0504045].

\bibitem{Seery:2005gb}
  D.~Seery and J.~E.~Lidsey,
  JCAP {\bf 0509}, 011 (2005)
  [arXiv:astro-ph/0506056].
  
\bibitem{Creminelli:2003iq}
  P.~Creminelli,
  JCAP {\bf 0310}, 003 (2003)
  [arXiv:astro-ph/0306122].
  


\bibitem{Seery:2005wm}
  D.~Seery and J.~E.~Lidsey,
  JCAP {\bf 0506}, 003 (2005)
  [arXiv:astro-ph/0503692].
  
  
\bibitem{Chen:2006nt}
  X.~Chen, M.~x.~Huang, S.~Kachru and G.~Shiu,
  JCAP {\bf 0701}, 002 (2007)
  [arXiv:hep-th/0605045].
  
 
  
\bibitem{Komatsu:2010fb}
  E.~Komatsu {\it et al.},
  arXiv:1001.4538 [astro-ph.CO].
  
\bibitem{Seery:2006vu}
  D.~Seery, J.~E.~Lidsey and M.~S.~Sloth,
  JCAP {\bf 0701}, 027 (2007)
  [arXiv:astro-ph/0610210].
  

\bibitem{Byrnes:2006vq}
  C.~T.~Byrnes, M.~Sasaki and D.~Wands,
  Phys.\ Rev.\  D {\bf 74}, 123519 (2006)
  [arXiv:astro-ph/0611075].
  
\bibitem{Smidt:2010sv}
  J.~Smidt, A.~Amblard, A.~Cooray, A.~Heavens, D.~Munshi and P.~Serra,
  arXiv:1001.5026 [astro-ph.CO].
  
\bibitem{ArmendarizPicon:1999rj}
  C.~Armendariz-Picon, T.~Damour and V.~F.~Mukhanov,
  Phys.\ Lett.\  B {\bf 458}, 209 (1999)
  [arXiv:hep-th/9904075].
 
\bibitem{Langlois:2008qf}
  D.~Langlois, S.~Renaux-Petel, D.~A.~Steer {\it et al.},
  Phys.\ Rev.\  {\bf D78}, 063523 (2008).
  [arXiv:0806.0336 [hep-th]].
  
 
 \bibitem{st}
  E.~Silverstein and D.~Tong,
  Phys.\ Rev.\  D {\bf 70}, 103505 (2004)
  [arXiv:hep-th/0310221];
  
 
  
\bibitem{ast}  
M.~Alishahiha, E.~Silverstein and D.~Tong,
  Phys.\ Rev.\  D {\bf 70}, 123505 (2004)
  [arXiv:hep-th/0404084].
    
 

\bibitem{Langlois:2009ej}
  D.~Langlois, S.~Renaux-Petel, D.~A.~Steer,
  JCAP {\bf 0904}, 021 (2009).
  [arXiv:0902.2941 [hep-th]].
  
  \bibitem{Gordon:2000hv}
  C.~Gordon, D.~Wands, B.~A.~Bassett and R.~Maartens,
  Phys.\ Rev.\ D {\bf 63}, 023506 (2001)
  [arXiv:astro-ph/0009131].

  
\bibitem{Langlois:2008wt}
  D.~Langlois, S.~Renaux-Petel, D.~A.~Steer {\it et al.},
  Phys.\ Rev.\ Lett.\  {\bf 101}, 061301 (2008).
  [arXiv:0804.3139 [hep-th]].
  
\bibitem{RenauxPetel:2009sj}
  S.~Renaux-Petel,
  JCAP {\bf 0910}, 012 (2009).
  [arXiv:0907.2476 [hep-th]].
  
\bibitem{Mizuno:2009mv}
  S.~Mizuno, F.~Arroja, K.~Koyama,
  Phys.\ Rev.\  {\bf D80}, 083517 (2009).
  [arXiv:0907.2439 [hep-th]].
 
  
\bibitem{Dvali:2003em}
  G.~Dvali, A.~Gruzinov and M.~Zaldarriaga,
  Phys.\ Rev.\  D {\bf 69}, 023505 (2004)
  [arXiv:astro-ph/0303591].
  
\bibitem{Kofman:2003nx}
  L.~Kofman,
  arXiv:astro-ph/0303614.
  





\bibitem{Langlois:2009jp}
  D.~Langlois and L.~Sorbo,
  JCAP {\bf 0908}, 014 (2009)
  [arXiv:0906.1813 [astro-ph.CO]].
  
 \bibitem{Chung:1999ve}
D.~J.~H.~Chung, E.~W.~Kolb, A.~Riotto and I.~I.~Tkachev,
   Phys.\ Rev.\  D {\bf 62}, 043508 (2000) [arXiv:hep-ph/9910437].
   
\bibitem{Elgaroy:2003hp}
  O.~Elgaroy, S.~Hannestad and T.~Haugboelle,
  JCAP {\bf 0309}, 008 (2003)
  [arXiv:astro-ph/0306229].

\bibitem{Romano:2008rr}
  A.~E.~Romano and M.~Sasaki,
  Phys.\ Rev.\  D {\bf 78}, 103522 (2008)
  [arXiv:0809.5142 [gr-qc]].
  
\bibitem{Barnaby:2009mc}
  N.~Barnaby, Z.~Huang, L.~Kofman {\it et al.},
  Phys.\ Rev.\  {\bf D80}, 043501 (2009).
  [arXiv:0902.0615 [hep-th]].
  
\bibitem{Barnaby:2010ke}
  N.~Barnaby,
  Phys.\ Rev.\  {\bf D82}, 106009 (2010).
  [arXiv:1006.4615 [astro-ph.CO]].
  
\bibitem{Suyama:2010uj}
  T.~Suyama, T.~Takahashi, M.~Yamaguchi {\it et al.},
  [arXiv:1009.1979 [astro-ph.CO]].

  \bibitem{curvaton}
  A.~D.~Linde and V.~F.~Mukhanov,
  Phys.\ Rev.\  D {\bf 56} (1997) 535
  [arXiv:astro-ph/9610219].
  K.~Enqvist and M.~S.~Sloth,
  Nucl.\ Phys.\  B {\bf 626}, 395 (2002)
  [arXiv:hep-ph/0109214];
  D.~H.~Lyth and D.~Wands,
  Phys.\ Lett.\  B {\bf 524}, 5 (2002)
  [arXiv:hep-ph/0110002];
  T.~Moroi and T.~Takahashi,
  Phys.\ Lett.\  B {\bf 522}, 215 (2001)
  [Erratum-ibid.\  B {\bf 539}, 303 (2002)]
  [arXiv:hep-ph/0110096].
 

\bibitem{Langlois:2004nn}
  D.~Langlois and F.~Vernizzi,
  Phys.\ Rev.\  D {\bf 70}, 063522 (2004)
  [arXiv:astro-ph/0403258];
  


\bibitem{Kawasaki:2008sn}
  M.~Kawasaki, K.~Nakayama, T.~Sekiguchi, T.~Suyama and F.~Takahashi,
  JCAP {\bf 0811}, 019 (2008)
  [arXiv:0808.0009 [astro-ph]].



\bibitem{Langlois:2008vk}
  D.~Langlois, F.~Vernizzi and D.~Wands,
  JCAP {\bf 0812}, 004 (2008)
  [arXiv:0809.4646 [astro-ph]].
    

\bibitem{Kawasaki:2008pa}
  M.~Kawasaki, K.~Nakayama, T.~Sekiguchi, T.~Suyama and F.~Takahashi,
  JCAP {\bf 0901}, 042 (2009)
  [arXiv:0810.0208 [astro-ph]].


\bibitem{Hikage:2008sk}
  C.~Hikage, K.~Koyama, T.~Matsubara, T.~Takahashi and M.~Yamaguchi,
  Mon.\ Not.\ Roy.\ Astron.\ Soc.\  {\bf 398}, 2188 (2009)
  [arXiv:0812.3500 [astro-ph]].



\bibitem{Kawakami:2009iu}
  E.~Kawakami, M.~Kawasaki, K.~Nakayama and F.~Takahashi,
  JCAP {\bf 0909}, 002 (2009)
  [arXiv:0905.1552 [astro-ph.CO]].
  
   
  
\bibitem{Langlois:2010dz}
  D.~Langlois, A.~Lepidi,
  JCAP {\bf 1101}, 008 (2011).
  [arXiv:1007.5498 [astro-ph.CO]].
  
\bibitem{Langlois:2010fe}
  D.~Langlois, T.~Takahashi,
  JCAP {\bf 1102}, 020 (2011).
  [arXiv:1012.4885 [astro-ph.CO]].
   

  
   
 



   
 


 
\bibitem{Moroi:2002rd}
  T.~Moroi and T.~Takahashi,
  Phys.\ Rev.\  D {\bf 66}, 063501 (2002)
  [arXiv:hep-ph/0206026].



\bibitem{Lyth:2003ip}
  D.~H.~Lyth and D.~Wands,
  Phys.\ Rev.\  D {\bf 68}, 103516 (2003)
  [arXiv:astro-ph/0306500].


  
   

   

\bibitem{Langlois:1999dw}
  D.~Langlois,
  Phys.\ Rev.\ D {\bf 59}, 123512 (1999)
  [arXiv:astro-ph/9906080].


 
\bibitem{Langlois:2000ar}
  D.~Langlois and A.~Riazuelo,
  Phys.\ Rev.\  D {\bf 62}, 043504 (2000)
  [arXiv:astro-ph/9912497].
  
  
  



 
\end{thebibliography}
\end{document}